# Isotopic separation during plasma expansion

**R.ANNOU**

**Faculty of Physics, USTHB, Algiers, ALGERIA.**

## Abstract

A plasma based isotopic separation method is proposed. Isotopes of diffferent masses get separated during plasma expansion. Relying on Gurevich's (*A.V.Gurevich, L.V.Pariï skaya and L.P.Pitaevskii, Sov.Phys.JETP, **36**,274 (1973)* ) plasma expansion into a vacuum model, the enrichment factor has been calculated. For ô =15 (ô being the normalized time), an increase of the relative abundance of 30% is expected.

In nature, uranium is found in a mixture state consisting of approximately 1% of $^{235}U$ and 99% of $^{238}U$ . Unfortunately, it is the least abundant isotope ($^{235}U$) that is of prime importance in many situations, e.g., in thermal neutrons reactors it is the isotope that sustains the chain reaction; hence, it is necessary to increase its relative abundance. This process being the enrichment. For research reactors, enrichment ranges from 6 to 90 %, whereas it ranges from 1.1 to 1.5 % for power reactors. To this end, several methods have been devised to achieve uranium enrichment, viz., gaseous diffusion, centrifugation, laser based and plasma based isotopic separation methods[1]. In this letter, we propose a method based on plasma peculiarities, viz., the separation of ions during plasma expansion into vacuum.

We consider a plasma of electrons and $^{238}U^{+}$ ions along with impurity ions $^{235}U^{+}$. We rely then on Gurevich *et al.*[2] model of a plasma expanding into vacuum, in order to derive the densities of the ions and the enrichment factor accordingly, with respect to the self-similar variable ô. The evolution of the ions distribution functions g are described by,

$$(u_1 - t)\frac{\partial g_1}{\partial t} - \frac{1}{2}\frac{\partial g_1}{\partial u_1}\frac{\partial y}{\partial t} = 0,$$

$$(u_2 - t)\frac{\partial g_2}{\partial t} - \frac{1}{2}\frac{\partial g_2}{\partial u_2}\frac{\partial y}{\partial t} = 0, \qquad (1)$$

where, $y(t) = \ln\left(\frac{1}{\sqrt{p}}\int g_1 d u_1\right)$, $u_j = V_{xj}/\sqrt{2T_e/M_j}$ and $t = x/t \times \sqrt{2T_e/M_j}$, with j=1,2 stand for $^{238}U^{+}$ and $^{235}U^{+}$ (the least abundant ion). All the parameters are having their usual significance.

At large values of ô, the thermal spread of the velocities may be neglected, due to the increase of the translational velocities.

Hence the following distributions,

$$g_1 = \sqrt{p}\, N_1(t)\, d\big(u_1 - u_1(t)\big),$$

$$g_2 = \sqrt{p}\, N_2(t)\, d\big(u_2 - u_2(t)\big), \qquad (2)$$

where, ä (x) is Dirac's delta function.

By virtue of Eq.(2), Eq.(1) may be cast as,

$$\left(u_1 - t\right)\frac{\partial N_1}{\partial t} + N_1\frac{\partial u_1}{\partial t} = 0\,,$$

$$\left(u_1 - t\right)\frac{\partial u_1}{\partial t} + \frac{1}{2}\frac{\partial y}{\partial t} = 0\,, \tag{3a}$$

and

$$\left(u_2 - t\right)\frac{\partial N_2}{\partial t} + N_2\frac{\partial u_2}{\partial t} = 0\,,$$

$$\left(u_2 - t\right)\frac{\partial u_2}{\partial t} + \frac{1}{2}\frac{M_1}{M_2}\frac{\partial y}{\partial t} = 0\,. \tag{3b}$$

For large values of ô , Eqs.(3a) and (3b) yield,

$$N_1(t) = \frac{N_{10}}{\sqrt{2}\,e\,z_1}e^{-t/z_1}\,,$$

$$N_2(t) = \frac{N_{20}}{\sqrt{2}\,e\,z_2}e^{-t/z_2}\,, \tag{4}$$

where, $z_1 = 1/\sqrt{2}$ and $z_2 = M_1/\sqrt{2}M_2$. $N_{10}$ and $N_{20}$ being the values of the densities at $t = -1/\sqrt{2}$, viz., the value of $t$ below which the plasma is unperturbed.

Using Eq.(4) one may calculate the relative abundance,

$$R = \frac{N_2}{N_1} = R_0 \exp\left(\frac{3\sqrt{2}}{238}t\right), \tag{5}$$

that leads to the enrichment factor ,

$$a^* = \frac{R}{R_0} = \exp\left(\frac{3\sqrt{2}}{238}t\right). \tag{6}$$

For ô =15, we find an increase of the relative abundance of 30%, whereas for ô ~ 36 an increase of 90% is expected.

In summary, a novel plasma based method of uranium enrichment is proposed. It is worthwile recalling that the plasma based methods are essentially confined to two major techniques, i.e., plasma centrifugation and electromagnetic wave-plasma interaction based isotopic separation. In the present case, i.e., during plasma expansion, ions of different masses get separated in space. The electrons move ahead under the influence of their own pressure, leading to the generation of a self-consistent electric field that drives the ions according to their respective inertia. Using a simplified plasma expansion model, we found an increase of 30% of the relative abundance for ô =15. It is clear however, that the present work gives the framework of a thought experiment only, hence in order to meet the requirement of a real situation, we have to cope with a set of physical parameters such as different reaction rates occuring during the process (e.g., recombination..). We emphasize also that isotopic separation is highly relevant in medical physics, especially that very small quantities are needed.

**References**

1. D. Massignon, *cycle du combustible: enrichissement de l'uranium*, Techniques de l'ingénieur, traité Mécanique et Chaleur, **B3600**.
2. A.V.Gurevich, L.V.Pariïskaya and L.P.Pitaevskii, Sov.Phys.JETP, **36**,274 (1973).